# Unoccupied space and short-range order characterization in polymers under heat treatment

Hossein Goodarzi Hosseinabadi [a] [*]

[a] Faculty of Engineering Sciences, Department of Biofabrication, University of Bayreuth, Ludwig Thoma Str. 36A, 95447 Bayreuth, Germany
hgoodarzy@gmail.com

**Abstract**

Large scale molecular dynamics simulations on polyvinyl alcohol were used to investigate the distribution of unoccupied space under different heat treatments. Representative volume elements consisting of 3600 chains of 300 monomers were equilibrated at melt state and cooled by different cooling rates. The positions of center of mass of the monomers were extracted and a series of spatial analysis were conducted to estimate the distribution of free volume or unoccupied space by Voronoi tessellation algorithm. An open-source software was employed which incorporates both compilers of Matlab and C programing languages to reduce the computational costs associated with Voronoi calculations and statistical analysis. The results confirmed that low free volume content is achievable through annealing while high free volume content is achievable through quenching samples at high cooling rates. An appreciable degree of short-range order in the packing of chains is revealed in annealed samples.

**Introduction**

Decades ago the concept of "free volume" was introduced to describe and formulate temperature dependent dynamics and mechanical properties in polymers [1–3]. However, the notion that the properties of the free space between monomers, known as free volume, is responsible for many aspects of polymeric behaviors is increasingly become controversial [4–6]. Experimental measurements of the unoccupied space between polymeric monomers for example using the annihilation of positrons are challenging and predictive while no exact estimation is yet available [5]. Therefore, the free volume estimations are tied to the used method for free volume characterization. Here, computational approaches based on molecular dynamics simulations has proven to be useful tools for quantitative estimation of the free volume properties. Three-dimensional Voronoi tessellation on the results of molecular dynamics simulations has shown to be a robust algorithm for analysis of the structure in polymers [7,8]. To this aim, after solving a molecular dynamics problem the positions of center of mass of the monomers are extracted to be used in Voronoi algorithm. The volumes of Voronoi polyhedrons are then recorded to estimate the distribution of free volume or unoccupied space. Rigby et al. [9] performed molecular dynamics on a system of 25 chains each having 20 monomers and measured the free volume distribution by Voronoi tessellation. They found that distribution of volumes of Voronoi polyhedral is broad at high temperatures but becomes narrower as the temperature is lowered toward the glass transition temperature. They reported that by decreasing the temperature of the polymer melt, an appreciable degree of short-range order in the packing of chains is observable. Performing the Voronoi analysis on large scale molecular systems takes high computational costs and resources. Here, the results of Voronoi tessellation and associated statistical analysis on a large-scale molecular dynamics model considering of $1.08\times10^6$ monomers is reported. In summary, a meaningful difference in the distribution of occupied volumes between polymeric monomers is observed by changing the cooling rate of the melt.

**Material and method**

The results obtained from a molecular dynamics simulation in a previous study [10] were used in current research. In brief, simulations were performed on a coarse-grained model representing polyvinyl alcohol (CG-PVA) as a semi-crystalline polymer. By changing the cooling rate, the degree of crystallinity, the mechanical properties and the spatial arrangement of monomers were remarkably affected [10,11]. Molecular dynamics simulations were performed using LAMMPS code considering 3600 chains of 300 monomers, the chain length. The bond length was considered 0.26 nm. Periodic boundary conditions were applied using a Berendsen barostat and a Nose-Hoover thermostat [12]. The time step used in all simulations was 0.00655 ps. The range and strength of 6-9 Lennard-Jones potential for non-bonded interactions were considered $\sigma_{LJ}= 0.89\times0.52$ and $\varepsilon_{LJ}= 1.511\times k_B T_0$ where $k_B$ is the Boltzmann constant and $T_0=550$ K is the reference temperature of the PVA melt [13]. The temperatures are reported in reduced unit $T= T_{real}/T_0$ and the time unit from the conversion relation of units is $\tau=1.31$ps. Equilibrated melts at T=1 were cooled with different cooling rates ($\dot{T}$) in the range $2\times10^{-7} \tau^{-1} <\dot{T}<10^{-3} \tau^{-1}$. Then the X, Y, Z positions of the centre of mass of each monomer in the RVE box was recorded as a seed to generate Voronoi polyhedral for free volume analysis [9]. The direction vector of each Voronoi domain consisting of $N_s$ facets was calculated by Eq. 1:

$$\vec{V}_P = \sum_{n=1}^{N_s} A_n (\vec{v}_x + \vec{v}_y + \vec{v}_z)\Big|_n = \sum_{n=1}^{N_s} A_n \sum_{i=1}^{3} \vec{v}_i^n \qquad (1)$$

where, $A_n$ is the area of facet 'n' and $\vec{v}_i^n$ is the vector to each facet vertices. The magnitude of the direction vector which represents the volume of Voronoi was determined by Eq. 2 [14].

$$|\vec{V}_p| = \sqrt{\langle \vec{V}_p \cdot \vec{V}_p \rangle} = \sqrt{(\vec{v}_x^p)^2 + (\vec{v}_y^p)^2 + (\vec{v}_z^p)^2} \qquad (2)$$

Then, the distribution of the free volume is roughly estimated by the distribution of volumes of Voronoi polyhedrons [9]. Although Voronoi tessellation does not give us the exact amount of free volume in the system, the size distribution of the Voronoi polyhedrons should be correlated to that of the free volume in the system [15]. The challenge is that running 3D Voronoi analysis for a large number of monomers takes a high computational cost. In current research, an open source software SpatialAnalysis3D™ is employed which incorporates both compilers of **Matlab** and **C** programing languages to accelerate calculations and reduce computational costs [14].

**Results and discussions**

Fig. 1-a shows histogram of free volume distribution obtained from different cooling rates when only 1200 monomers at the middle of the simulation box cooled from melting temperature are considered for the 3D

Voronoi analysis. Many fluctuations are seeable in the obtained histograms showing that the number of monomers used in Voronoi analysis were not sufficient to reach a homogenous distribution profile. By increasing the number of monomers for Voronoi analysis an extraordinary random-access memory (RAM) and computational time is required for calculating the volume of Voronoi polyhedrons. To provide a better estimation about the required computational resources for large scale Voronoi analysis, the analysis was repeated considering the whole molecular dynamics model at Discovery platform in the Massachusetts Green High Performance Computing Centre (MCHPC). Fig. 1-b represents the actual RAM and computational time requirements for calculating the free volume distribution with contribution of larger number of monomers. Time and RAM both increase with a parabolic function of representative volume element (RVE) or the number of monomers. To run the Voronoi analysis for 1'080'000 monomers, a non-conventionally accessible RAM as high as 8 terabytes is required.

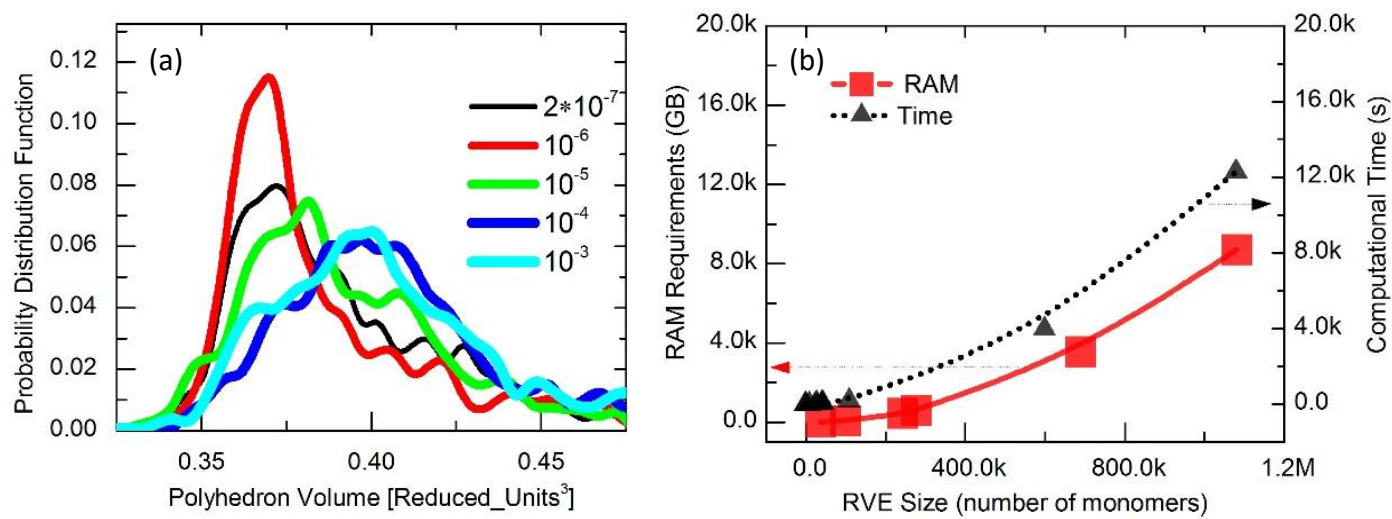

Fig. 1- a) occupied volume distribution for different cooling rates for 1200 monomers; b) Computational time and RAM requirements for Voronoi analysis using different number of monomers.

Fig. 2 illustrates the sensitivity analysis to the number of monomers used for Voronoi calculations. It is shown that considering equal to or more than 30'000 monomers is sufficient for obtaining reproducible results.

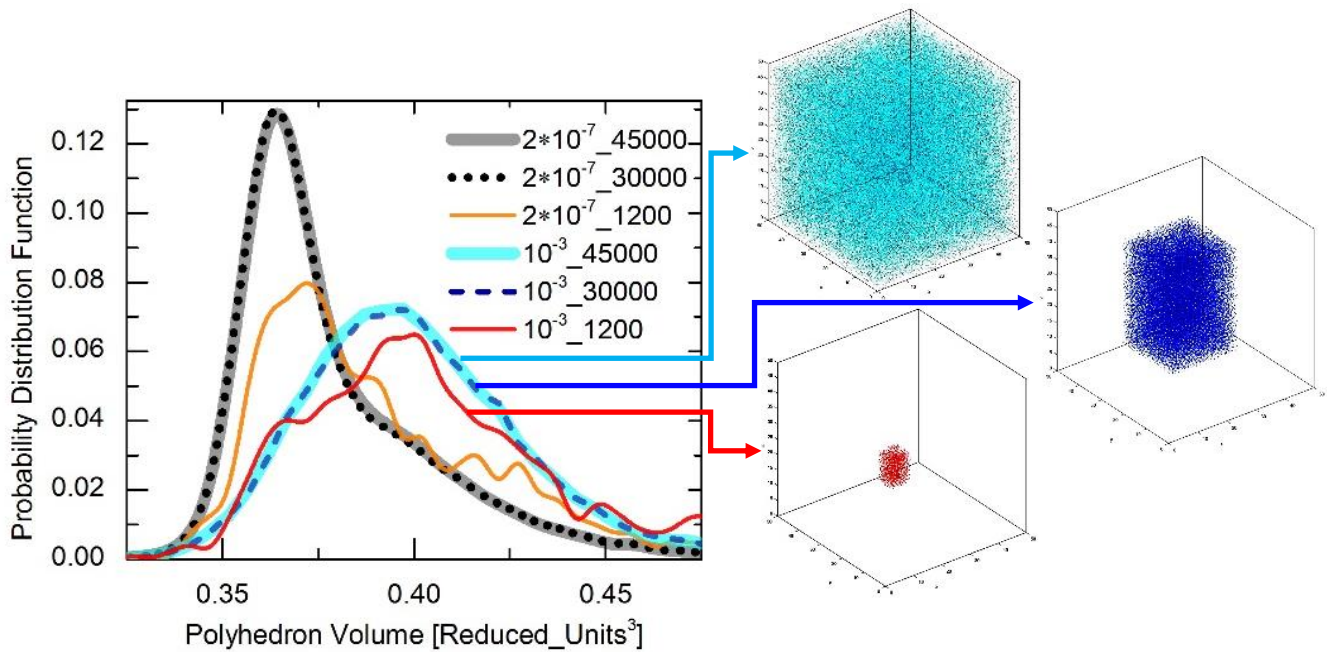

Fig. 2- Sensitivity of Voronoi analysis to the number of monomers used for Voronoi analysis in two different cooling rates.

In molecular dynamics simulations under periodic boundary conditions, monomer positions are typically wrapped into a reference box. For some applications such as diffusion coefficient calculations using the Einstein relation for example, the particle positions need to be unwrapped [16]. Fig. 3 compares the free volume analysis for both wrapped and unwrapped situations at two different cooling rates. It is evident that the Voronoi analysis on the central part of RVE returns similar results in both wrapped and unwrapped configurations. However, the analysis on boundary/edge sections in unwrapped configuration involves higher noise, scattering, and larger average volume in Voronoi algorithm.

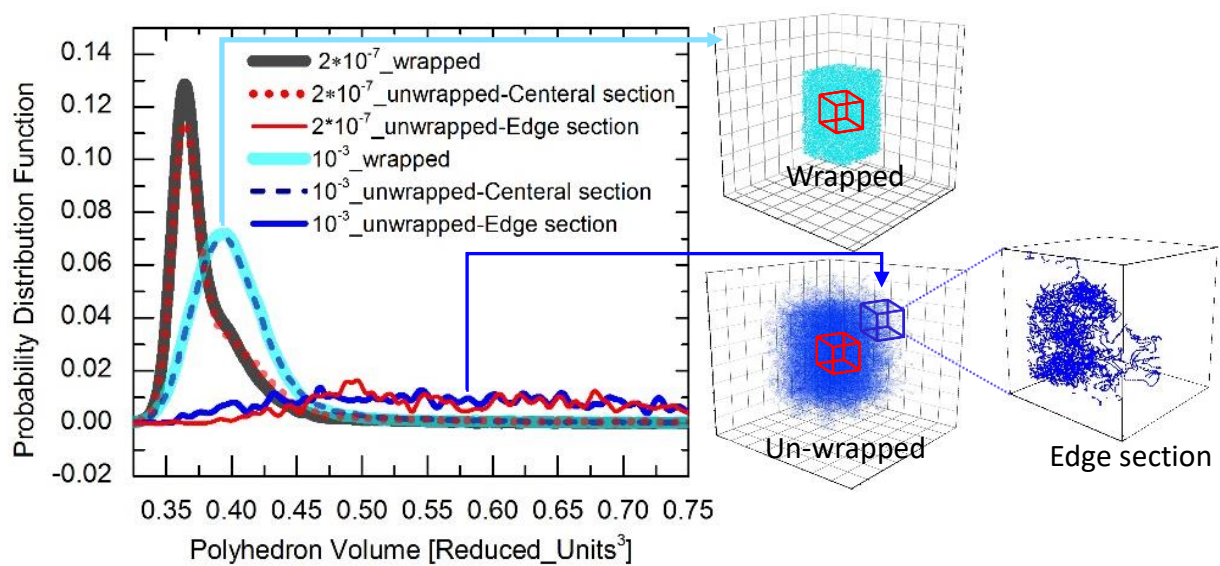

Fig. 3- Voronoi analysis on different parts of the RVE under wrapped/unwrapped configurations.

Figure 4 shows that the free volume or the volume of a Voronoi polyhedron for a chain end (side-monomers) is larger than that for an internal monomer in all examined cooling rates. This agrees well with previous reports obtained from united-atom molecular dynamics simulations of polyethylene [17].

In addition, distribution of volumes of Voronoi polyhedral for internal monomers is broad when high cooling rate ($10^{-3}$) is applied while it becomes narrower when low cooling rate ($10^{-7}$) is applied. Indeed, under high cooling rates the monomers conformation freeze from the melt state where a larger size and wider distribution of free volume is reported [18].

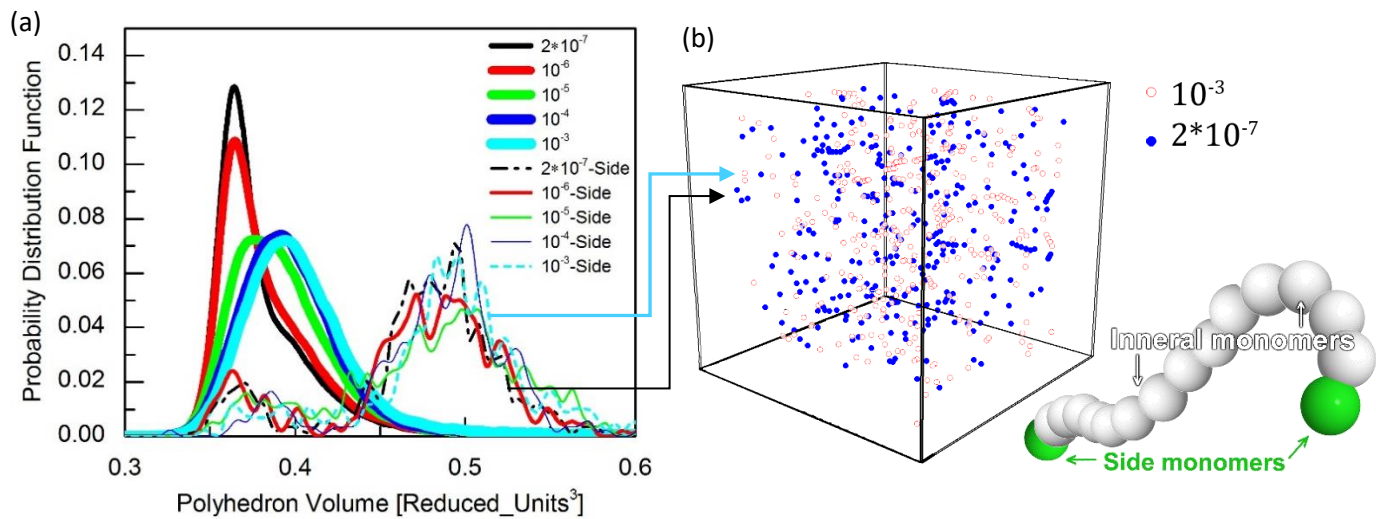

Fig. 4- Distribution of volume of polyhedrons formed at internal and end/side monomers of each chain, with visual demonstration of end/side monomers for the lowest ($10^{-7}$) and highest ($10^{-3}$) cooling rates.

Fig. 5, illustrates the free volume distribution for different cooling rates as well as the spatial arrangement of the monomers for the lowest and highest cooling rates. A short-range order in arrangement of monomers at the lowest cooling rate (annealing at $10^{-7}$) is evident. In other words, by slow decreasing of the temperature of the polymer melt, an appreciable degree of short-range order in the packing of the chains is observable.

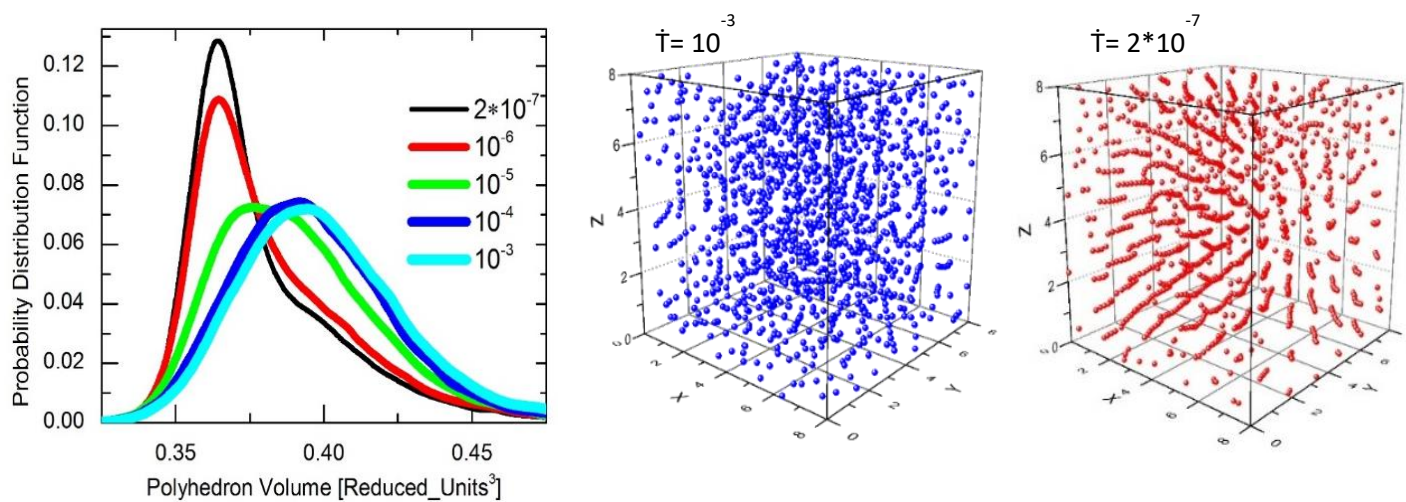

Fig. 5- Histograms of Voronoi polyhedrons to characterize the distribution of the free volume; insets can be added to show the related microstructural orientations in 3X magnification for the lowest ($10^{-7}$) and highest ($10^{-3}$) cooling rates.

Fig. 6-b visualizes the projection of Voronoi direction vectors obtained from Eq. 2 in 2D space for different cooling rates. Higher randomness in projected direction vectors is evident in patterns obtained from higher cooling rates (quenched samples), while clusters of direction vector (marked by arrows in the figure) can be seen in low cooling rates (annealed samples) which may represent the crystallinity in the structure. Fig. 6-c, represents a quantitative evaluation of the components of Voronoi direction vectors in X, Y, and Z directions. It is evident that by decreasing the cooling rate, the alignment of direction vectors is improved and the directionality of the system is increased by increasing the crystallinity in annealed samples, in accordance with previous reports on similar systems [11,12].

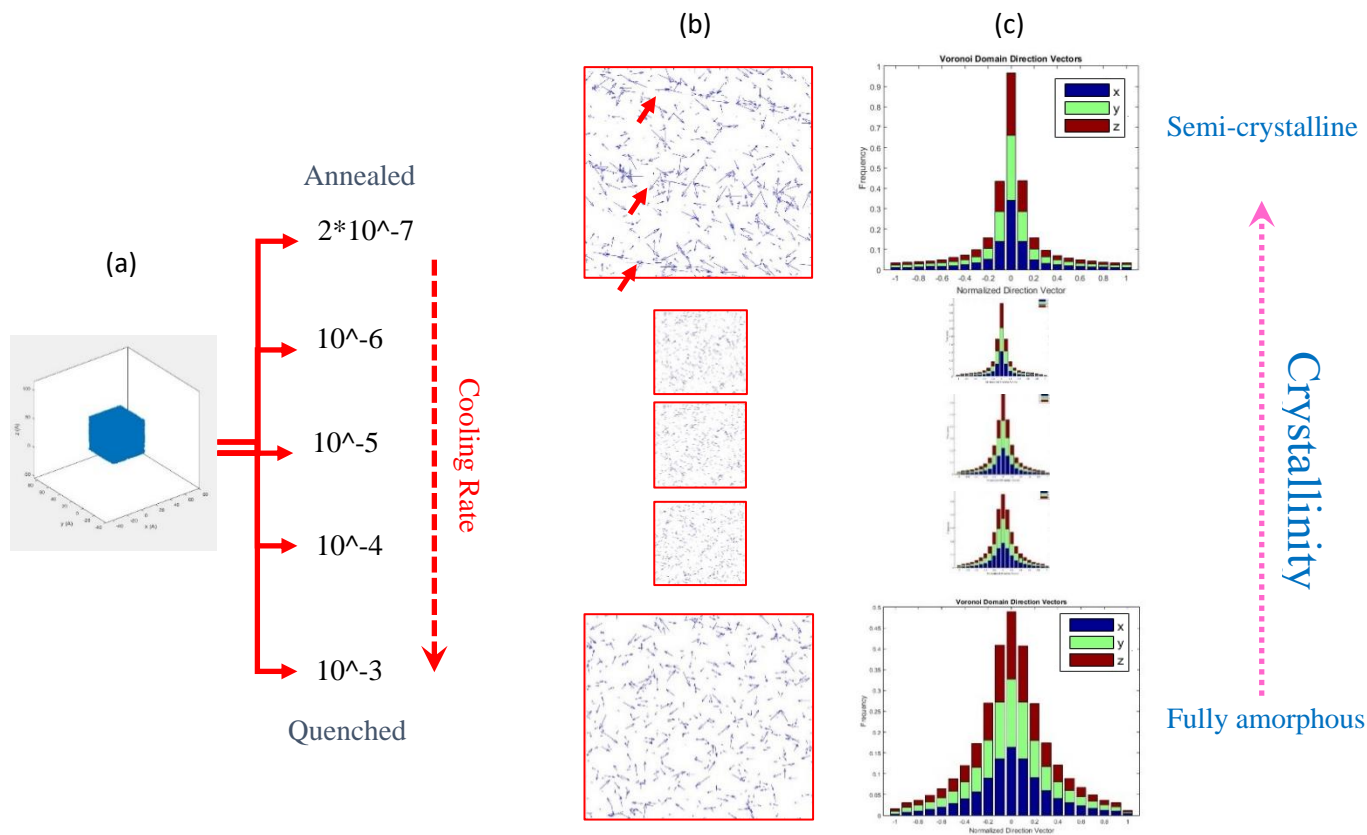

Fig. 6- Characterization of free volume and visualization of spatial arrangement of monomers by analyzing MD results, a) Isometric view of representative volume element, b) 6X magnification (XY plane) to visualize Voronoi direction vectors in 2D, c) quantitative evaluation of components of Voronoi direction vectors in X, Y, Z (see figure S1)

**Conclusion**

Results of large-scale molecular dynamics simulations were used to investigate the distribution of unoccupied space using Voronoi tessellation. We showed that the Voronoi analysis allows to characterize the free volume or unoccupied space at length scales smaller than 100 nm which is not feasible by experiment. The analysis revealed that the lowest free volume content is achievable through annealing while the highest free volume content is achievable through quenching the melt at high cooling rates. An appreciable degree of short-range order in the packing of chains was observed in annealed samples.

**Acknowledgement**

Thanks to Dr. S. Jabbari-Farouji for providing the LAMMPS outputs and useful discussions. The author is thankful for the support of AvH foundation which in part contributed to concluding this research. Performing the computations presented in this research was initiated at Northeastern University using the Discovery platform, which was a part of the high-performance computing resource in the Massachusetts Green High Performance Computing Center.

*References*
1. Kwag, C., Manke, C. W. & Gulari, E. Effects of Dissolved Gas on Viscoelastic Scaling and Glass Transition Temperature of Polystyrene Melts. *Ind. Eng. Chem. Res.* **40**, 3048–3052 (2001).
2. Hölck, O., Böhning, M., Heuchel, M., Siegert, M. R. & Hofmann, D. Gas sorption isotherms in swelling glassy polymer - detailed atomistic simulations. *J. Memb. Sci.* **428**, 523–532 (2013).
3. Van Melick, H. G. H., Govaert, L. E., Raas, B., Nauta, W. J. & Meijer, H. E. H. Kinetics of ageing and re-embrittlement of mechanically rejuvenated polystyrene. *Polymer (Guildf).* **44**, 1171–1179 (2003).
4. White, R. P. & Lipson, J. E. G. Polymer Free Volume and Its Connection to the Glass Transition. *Macromolecules* **49**, 3987–4007 (2016).
5. Hosseinabadi, H. G. *et al.* Interrelation between mechanical response , strain field , and local free volume evolution in glassy polymers : Seeking the atomistic origin of post-yield softening. *Express Polym. Lett.* **12**, 2–12 (2018).
6. Van-Melick, H. G. H., Govaert, L. E. & Meijer, H. E. H. On the origin of strain hardening in glassy polymers. *Polymer (Guildf).* **44**, 2493–2502 (2003).
7. He, B. *et al.* CAVD, towards better characterization of void space for ionic transport analysis. *Sci. Data 2020 71* **7**, 1–13 (2020).
8. Sung, B. J. & Yethiraj, A. Structure of void space in polymer solutions. *Phys. Rev. E* **81**, 031801 (2010).
9. Rigby, D. & Roe, R. J. Molecular Dynamics Simulation of Polymer Liquid and Glass . 4 . Free-Volume Distribution. *Macromolecules* **23**, 5312–5319 (1990).
10. Kryuchkov, N. P., Yurchenko, S. O., Fomin, Y. D., Tsiok, E. N. & Ryzhov, V. N. Complex crystalline structures in a two-dimensional core-softened system. *Soft Matter* **14**, 2152–2162 (2018).
11. Jabbari-Farouji, S. *et al.* Plastic Deformation Mechanisms of Semicrystalline and Amorphous Polymers. *ACS Macro Lett.* **4**, 147–150 (2015).
12. Jabbari-Farouji, S. *et al.* Correlation of structure and mechanical response in solid-like polymers. *J. Phys. Condens. Matter* **27**, 194131 (2015).
13. Meyer, H. & Müller-Plathe, F. Formation of chain-folded structures in supercooled polymer melts. *J. Chem. Phys.* **115**, 7807 (2001).
14. Dan, L. *Spatial Analysis 3D' : Statistical and visualization program for three-dimensional spatial point patterns*. *Analysis* (University of California at Santa Barbara, 2007).
15. Wong, C. P. J. & Choi, P. On the diffusivity of ring polymers. *Soft Matter* **16**, 2350–2362 (2020).
16. Bülow, S. von, Bullerjahn, J. T. & Hummer, G. Systematic errors in diffusion coefficients from long-time molecular dynamics simulations at constant pressure. *J. Chem. Phys.* **153**, 021101 (2020).
17. Tokita, N., Hirabayashi, M., Azuma, C. & Dotera, T. Voronoi space division of a polymer:

Topological effects, free volume, and surface end segregation. *Journal of Chemical Physics* **120**, 496–505 (2004).
18. Heuchel, M., Fritsch, D., Budd, P. M., McKeown, N. B. & Hofmann, D. Atomistic packing model and free volume distribution of a polymer with intrinsic microporosity (PIM-1). *J. Memb. Sci.* **318**, 84–99 (2008).